\documentclass[fleqn,10pt]{wlscirep}
\title{Ultrafast Imaging of Laser Driven Shock Waves using Betatron X-rays from a Laser Wakefield Accelerator}

\usepackage{siunitx}
\usepackage[colorinlistoftodos]{todonotes}

\usepackage{color}

\author[1]{\normalsize J.~C.~Wood}
\author[2]{\normalsize D.~J.~Chapman}
\author[1]{\normalsize K.~Poder}
\author[1,3]{\normalsize N.~C.~Lopes}
\author[2,4]{\normalsize M.~E.~Rutherford}
\author[5]{\normalsize T.~G.~White}
\author[6]{\normalsize F.~Albert}
\author[7]{\normalsize K.~T.~Behm}
\author[8]{\normalsize N.~Booth}
\author[1]{\normalsize J.~S.~J.~Bryant}
\author[8]{\normalsize P.~S.~Foster}
\author[9]{\normalsize S.~Glenzer}
\author[10]{\normalsize E.~Hill}
\author[7]{\normalsize K.~Krushelnick}
\author[1]{\normalsize Z.~Najmudin}
\author[6]{\normalsize B.~B.~Pollock}
\author[10]{\normalsize S.~Rose}
\author[9]{\normalsize W.~Schumaker}
\author[8]{\normalsize R.~H.~H.~Scott}
\author[10]{\normalsize M.~Sherlock}
\author[7]{\normalsize A.~G.~R.~Thomas}
\author[7]{\normalsize Z.~Zhao}
\author[2,4]{\normalsize D.~Eakins}
\author[1]{\normalsize S.~P.~D.~Mangles}

\affil[1]{\normalsize  \it The John Adams Institute for Accelerator Science, Blackett Laboratory, Imperial College London, London, SW7 2AZ, UK}
\affil[2]{\normalsize \it Institute of Shock Physics, Blackett Laboratory, Imperial College London, London, SW7 2AZ, UK}
\affil[3]{\normalsize \it GoLP/Instituto de Plasmas e Fus\~{a}o Nuclear, Instituto Superior T\'{e}cnico, U.L., Lisboa 1049-001, Portugal}
\affil[4]{\normalsize \it Solid Mechanics and Materials Engineering, Department of Engineering Science, University of Oxford, Oxford, OX5 1PF, UK}
\affil[5]{\normalsize \it Department of Physics, University of Nevada, Reno, Nevada 89557, USA}
\affil[6]{\normalsize \it Lawrence Livermore National Laboratory, 7000 East Ave, Livermore, CA 94550, USA}
\affil[7]{\normalsize \it Center for Ultrafast Optical Science, University of Michigan, 2200 Bonisteel Blvd., Ann Arbor, MI 48109-2099, USA}
\affil[8]{\normalsize \it Central Laser Facility, Rutherford Appleton Laboratory, Didcot, OX11 0QX, UK}
\affil[9]{\normalsize \it SLAC, 2575 Sand Hill Rd, Menlo Park, CA 94025, USA}
\affil[10]{\normalsize \it Blackett Laboratory, Imperial College London, London, SW7 2AZ, UK}

\affil[*]{jonathan.wood08@imperial.ac.uk}

\begin{abstract}
Betatron radiation from laser wakefield accelerators is an ultrashort pulsed source of hard, synchrotron-like x-ray radiation. 
It emanates from a centimetre scale plasma accelerator producing GeV level electron beams.
In recent years betatron radiation has been developed as a unique source capable of producing high resolution x-ray images in compact geometries. 
However, until now, the short pulse nature of this radiation has not been exploited. 
This report details the first experiment to utilise betatron radiation to image a rapidly evolving phenomenon by using it to radiograph a laser driven shock wave in a silicon target.
The spatial resolution of the image is comparable to what has been achieved in similar experiments at conventional synchrotron light sources.
The intrinsic temporal resolution of betatron radiation is below 100\,fs, indicating that significantly faster processes could be probed in future without compromising spatial resolution.
Quantitative measurements of the shock velocity and material density were made from the radiographs recorded during shock compression and were consistent with the established shock response of silicon, as determined with traditional velocimetry approaches.
This suggests that future compact betatron imaging beamlines could be useful in the imaging and diagnosis of high-energy-density physics experiments.

\end{abstract}
\begin{document}

\flushbottom
\maketitle

\thispagestyle{empty}

\section*{Introduction}

Understanding the behaviour of materials in extreme states of pressure and temperature is fundamental to multiple fields of study, from condensed matter to meteorite impacts \cite{Melosh1984,Trepmann2008}, behaviour in planetary cores \cite{Kanatzidis2000}, or material failure mechanisms \cite{Hudspeth2015a,Ramos2014}.
The only way to access states at very high energy densities in the laboratory are dynamic compression experiments, where a shock wave is driven into a material by, for example, a pulsed laser or a high velocity flyer plate.
These states are necessarily short lived because of the limited amount of energy that is available to sustain and support them. 
Traditional surface based measurements used to diagnose these experiments such as high speed photography, or surface velocity diagnostics such as VISAR \cite{Barker1972}, have the disadvantage that subsurface processes cannot be diagnosed directly. 
Without subsurface probing any measurement of the shock properties will be affected by unloading effects at free surfaces or buried material interfaces. 
Furthermore, fast, high resolution subsurface probing is required for a real-time, non-intrusive diagnosis of material behaviour at the microscale and is the only way to fully observe the effects of material inhomogeneities.
This has driven the use of bright, short-pulse x-ray sources as a diagnostic of rapidly evolving phenomena inside otherwise opaque materials \cite{Jensen2012a,Jensen2013}, which can be used in conjunction with velocimetry techniques to simultaneously probe surface and sub-surface dynamics.\\
To provide sufficient imaging contrast between features in solid density, moderate atomic number $\left( Z \approx 10-20 \right)$ materials, the x-ray source must be `hard' (10's\,keV). 
In order to temporally resolve fast physical processes it is desirable to have an ultrashort pulse duration.
To take high signal-to-noise ratio images, the source must have a large number of photons per unit solid angle with preferably micrometre-scale imaging resolution. 
To produce an x-ray pulse with all of these characteristics, light sources based on energetic, short pulse, low emittance electron beams are generally required, namely synchrotron light sources or more recently x-ray free electron lasers (XFELs). 
Synchrotron light sources produce monochromatic or broadband pulses of duration 10s to 100 picoseconds \cite{Jensen2012a} with peak brightness of order $ 10^{23} $\,photons/s/mm$^2$/mrad$^2/$0.1\%BW \cite{Ackermann2007}. 
XFELs produce monochromatic pulses with a duration of a few 10s femtoseconds with a brightness of $10^{30} - 10^{34}$\,photons/s/mm$^2$/mrad$^2/$0.1\%BW \cite{Fletcher2015a}. \\
Recently much work has been carried out to develop laser wakefield accelerators (LWFAs) that are capable of producing GeV \cite{Leemans2006a,Kneip2009,Clayton2010,Kim2013,Wang2013,Leemans2014}, few femtosecond \cite{Lundh2011}, low emittance (of order $1 \pi$\,mm-mrad) \cite{Brunetti2010} electron beams from centimetre scale plasma sources driven by high intensity optical lasers. 
In a LWFA a short, relativistically intense laser pulse, typically with intensity $ I \sim 10^{18} - 10^{19} $\,\si{\watt\per\square\centi\metre}, excites a nonlinear plasma wave that trails the laser pulse, which for such high intensities takes the form of an almost spherical bubble that is devoid of electrons \cite{Pukhov2002a,Lu2006a}. 
The accelerating fields within the bubble are of order 100\,\si{\giga\volt\per\metre}. 
The bubble also acts as a wiggler for the electrons that are being accelerated, resulting in a hard, bright, broadband source of so-called betatron x-rays \cite{Rousse2004a,Kneip2010a} with a pulse length of a few 10s femtoseconds \cite{Mangles2006a} (inferred from the electron bunch length), which is comparable to XFEL sources.
The flux of betatron x-rays is high enough for single pulse imaging using standard detectors such as a commercially available caesium iodide scintillator coupled to a scientific CCD camera, or Imaging Plate. 
The x-rays are emitted from a micrometre scale source \cite{Phuoc2006,Kneip2010a} which permits high resolution imaging in a compact, few metre long geometry, comparing favourably with the $\sim 100$\,m long imaging beamlines of conventional undulator and wiggler sources.
The small source size means that betatron radiation can be used for propagation based phase contrast imaging \cite{Kneip2011a,Fourmaux2011a,Wenz2015}, where image contrast is formed by the refraction of light from gradients in the real part of the refractive index in the target.
The betatron source has previously been reported to have a peak brightness of order $10^{22} - 10^{23}$\,photons/s/mm$^2$/mrad$^2/$0.1\%BW \cite{Kneip2010a,Cole2015,Yan2014}, comparable to that which is achieved at modern short pulse synchrotron beamlines. 
Betatron sources are compact, versatile and can thus be coupled to many different drivers in a variety of geometries. 
Their ultrashort, broad bandwidth x-ray pulses make them complementary to XFELs and are suitable for x-ray imaging and spectroscopic studies.
Additionally they could be implemented at existing large scale HEDP facilities with a much smaller capital outlay and footprint than would be required to install a synchrotron or XFEL source.\\
Here we present results from a proof of principle experiment where betatron radiation was used to image laser driven shock waves in silicon, demonstrating that this source can be used to temporally resolve dynamic phenomena that would previously have required a considerably larger and more expensive synchrotron machine. 
Indeed similar shock imaging experiments have recently been performed at the LCLS XFEL \cite{Schropp2015}, as well as multiple synchrotron facilities \cite{Jensen2012a,Rutherford2017}. 
It is shown in this Report that the LWFA betatron source can be used to take high-resolution x-ray images of shock waves propagating at multi-\si{\kilo\meter\per\second} velocities through a solid density silicon sample.
The measured shock velocity and density are in agreement with the silicon Hugoniot data and radiation hydrodynamics simulations.
Additionally it is demonstrated that phase contrast effects arising from the small source allow for simultaneous imaging of the deformation of CH ablator layers and the dynamics of higher-$Z$ shock targets.

\section*{Results}

\subsection*{Electron and X-ray Beam Characterisation}

\begin{figure}[h!]
\centering
\includegraphics[width=0.9\linewidth]{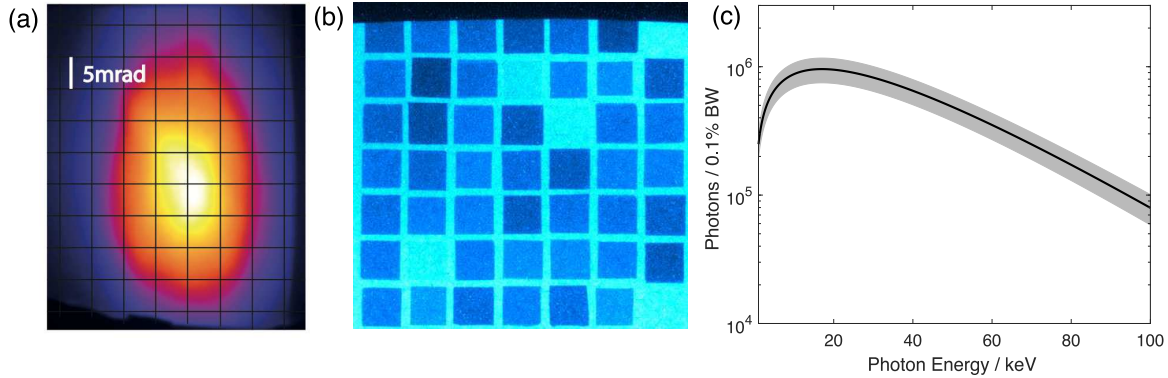}
\caption{X-ray characterisation results. (a)~Example image of the x-ray beam profile, where the vertical direction in the image is parallel to the laser polarisation direction. (b)~X-ray image of the filter array used to estimate the critical energy $E_c$. (c)~Mean number of photons per 0.1\% bandwidth in the whole beam where the error, shown by the grey band, is a combination of the uncertainties in the total photon number and $E_c$.}
\label{fig:photonspec}
\end{figure}

The experiment was performed at the dual beam Gemini laser system at the Central Laser Facility, Rutherford Appleton Laboratory, U.K. 
One arm of the 200\,TW Gemini laser was focussed in to a helium filled gas cell to drive a self-guided, self-injecting LWFA (see Methods). 
The accelerated electron beams had peak energies in the range $0.6-1.0$\,GeV, with 70\% of shots having a peak energy in the range $\left( 700 \pm 100 \right)$\,MeV. 
The total beam charge above 200\,MeV was $200-300$\,pC per shot. \\
The FWHM x-ray beam divergence was $\left(20 \pm 2 \right)\,\textrm{mrad} \times \left(8.9 \pm 0.7 \right)\,\textrm{mrad}$ in the directions parallel and perpendicular to the laser polarisation direction respectively.
An example image of the x-ray beam profile screen is shown in Figure \ref{fig:photonspec}~(a).
The x-ray spectrum was assumed to be an on-axis synchrotron spectrum \cite{Esarey2002,Fourmaux2011}, which for a single electron of energy $\gamma m_e c^2$ is given by
\begin{equation}
\frac{\textrm{d}^{2} I}{\textrm{d}E\textrm{d} \Omega} = \frac{3e^{2}}{16 \pi ^{3} \hbar c \epsilon _0} \gamma ^{2} \left(\frac{E}{E_{c}} \right) ^{2} K _{2/3} ^{2} \left(E/ 2E_{c} \right),
\label{eq:betatronspectrum}
\end{equation}
where $K _{2/3}$ is a modified Bessel function of the second kind and $E$ is the photon energy. 
The shape of the spectrum is determined by the critical energy $E_c$, which using a definition consistent with Jackson 3\textsuperscript{rd} edition \cite{jackson_classical_1999} is given by
\begin{equation}
E_c = \frac{3\hbar}{4c} \gamma^{2} \, {\omega_p}^2 \, r_{\beta},
\label{eq:Ecrit}
\end{equation}
where $\omega_p$ is the plasma frequency and $r_{\beta}$ is the oscillation radius of the electron.
The mean critical energy measured over 40 shots was $E_c = \left(20 \pm 5 \right)$\,keV, where the error represents the standard deviation.
$E_c$ was determined by comparing the transmission through an array of different elemental filters \cite{Kneip2008} (see Methods).
An example betatron x-ray image of the filter array is shown in Figure \ref{fig:photonspec}~(b).
Combining the measured $E_c$ with the peak electron energy and the plasma density in Equation \ref{eq:Ecrit} gives an estimated source size $2 r_{\beta} = (1.6 \pm 0.4)$\,\si{\um}. 
The mean number of photons per shot in the whole beam with $E>1$\,keV was $(5.5\pm 1.2) \times 10^9$, corresponding to a peak of $\left(1.4 \pm 0.3 \right) \times 10^6$\,photons\,/\,mm\textsuperscript{2} on the detector.
The number of photons per 0.1\% bandwidth inferred from these measurements is shown in Figure \ref{fig:photonspec}~(c).
Assuming that the betatron pulse duration $\tau_{\beta} =  60$\,fs, corresponding to a pulse length of approximately half of the nonlinear plasma wavelength i.e.~$c\tau_{\beta} \approx \sqrt{a_0} \lambda_p/2$, the peak brightness was $\left(6 \pm 1 \right) \times 10^{22}$\,photons/s/mm\textsuperscript{2}/mrad\textsuperscript{2}/0.1\%BW.\\
Images of the edge of a cleaved silicon crystal taken with a geometric magnification of 30 revealed that the FWHM of the line spread function (LSF) of the imaging system was $\left(140 \pm 10 \right)$\,\si{\um} at the plane of the scintillator, or $\left(4.7 \pm 0.4 \right)$\,\si{\um} at the target plane in this experiment. 
This demonstrated that few micrometre sized features could be resolved, which is comparable to similar experiments at synchrotron light sources \cite{Ramos2014,Jensen2012a,Yeager2012}. 
The target was placed 0.12\,m from the source and the camera was 3.70\,m from the source. 
A 1.6\,\si{\um} source, as measured, would produce a 46\,\si{\um} wide LSF on the detector, meaning that the resolution achievable in this geometry was mainly limited by the detector and not the betatron source. 

\subsection*{Betatron X-ray Imaging of Shocked Silicon}
\begin{figure}[h!]
\centering
\includegraphics[width=\linewidth]{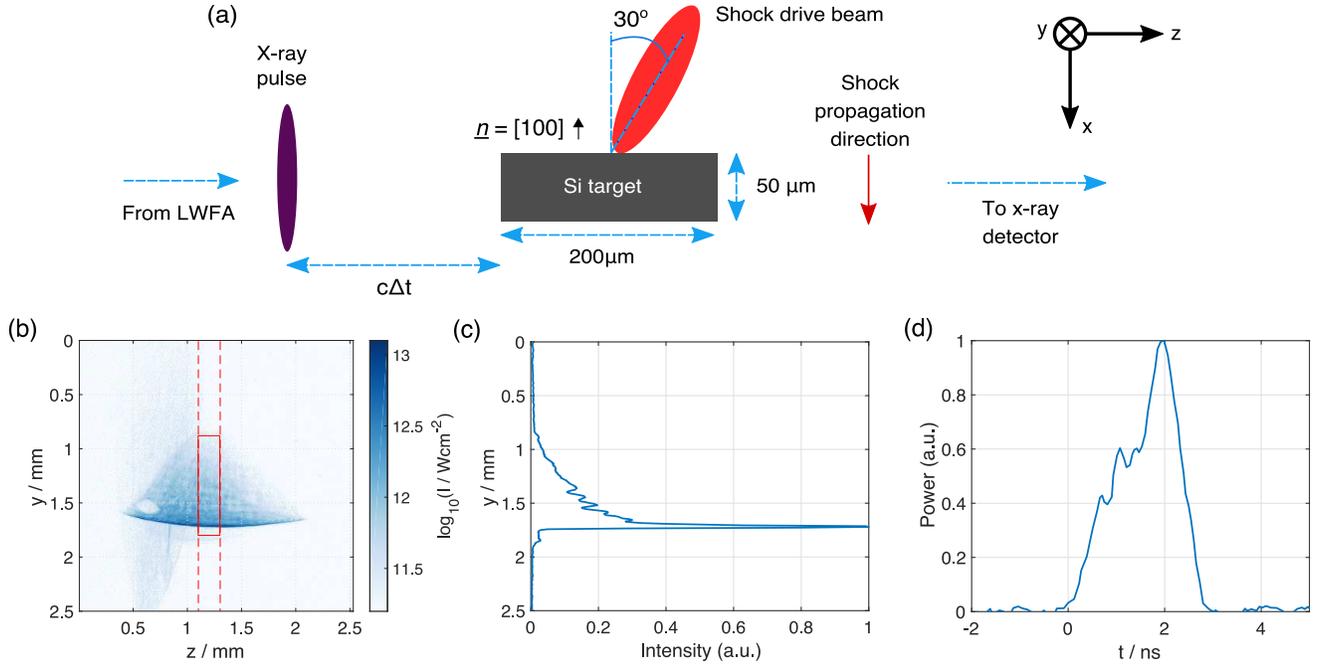}
\caption{Details of the interaction. (a)~Top view of the shock target interaction point. The shock drive beam arrived at $t=0$, and the target was probed by the ultrafast betatron x-ray beam at time $\Delta t$. 
The unit vector \underline{\textit{n}} indicates the crystal orientation. 
(b)~The laser intensity profile used to drive the shock, where a logarithmic scale has been used to highlight the spatial extent of the spot. The target dimensions are indicated by the red rectangle.
(c)~The laser intensity profile integrated along $z$ between the limits highlighted by the red dashed lines in part (b).
(d)~The temporal profile of the shock drive pulse as measured by a diode.}
\label{fig:shockimsetup}
\end{figure}

\begin{figure}[h!]
\centering
\includegraphics[width=0.9\linewidth]{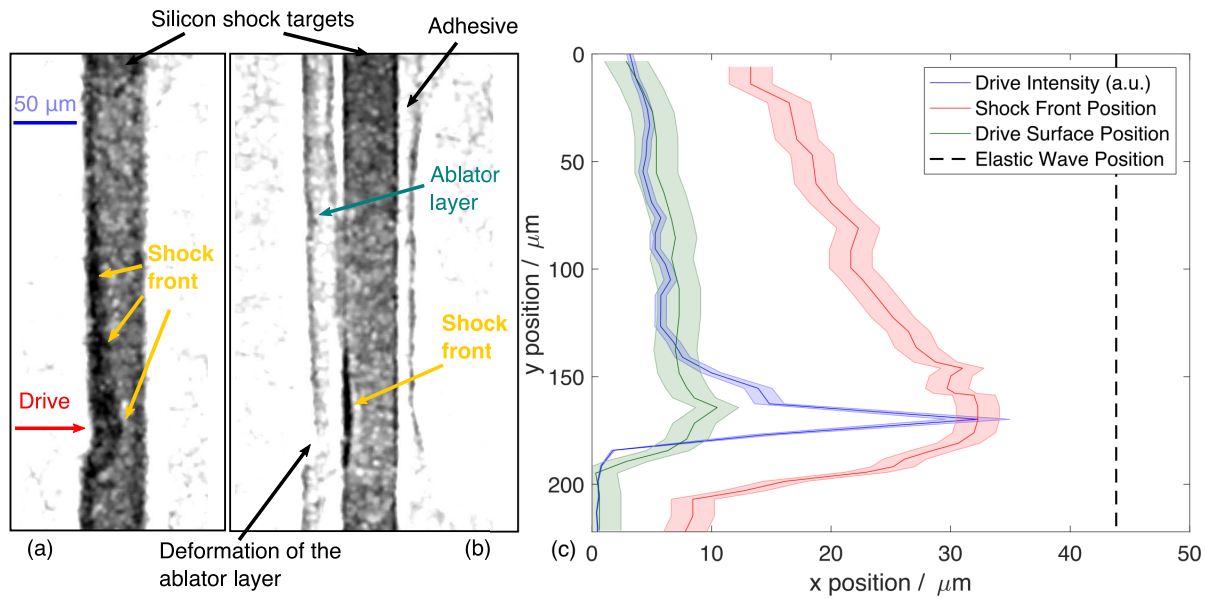}
\caption{X-ray images of shocked silicon targets taken with betatron radiation. (a)~Radiograph of a laser driven shock in silicon at 5.2\,ns after the start of the interaction.
The shock driving laser travelled from left to right. 
The geometric magnification of the images was 30.
(b)~Radiograph of a shock wave in silicon with a 25\,\si{\um} CH ablator layer on the drive surface taken at $\Delta t = 6.5$\,ns. 
Also visible is adhesive at the rear of the target, which did not participate in the interaction. 
(c)~Drive laser intensity (blue), shock front position (red) and drive surface position (green) as a function of $y$ found from the image of the untamped silicon sample, where the shaded area indicates the error. 
The black dashed line shows the position that the elastic wave, which was not observed in this experiment, would have reached after 5.2\,ns (travelling at $8.43$\,\si{\kilo\metre\per\second})\cite{McSkimin1964}.}
\label{fig:shockprop1}
\end{figure}

The geometry of the shock target interaction point is shown in Figure \ref{fig:shockimsetup}~(a). 
Shock waves were driven into a single crystal silicon target along the [100] direction. 
The drive laser pulse arrived at an angle of $30^{\circ}$ to this, so that reflected laser light would not be collected by the focussing optic and sent back along the laser chain. 
The silicon was 50\,\si{\um} thick in the shock propagation direction and 200\,\si{\um} thick in the betatron probe direction. 
200\,\si{\um} thick silicon is 50\% transmissive at a photon energy of 13.25\,keV, highlighting the need for a hard x-ray source.
The effect of having a 25\,\si{\um} thick plastic CH ablator layer on the shock driver side of the target was also explored. 
Ablator layers are commonly used to mitigate spatial heterogeneities in the drive laser focal spot thereby delivering a more spatially uniform shock, as well as helping to support the shock state for longer and prevent unwanted stochastic heating of the sample \cite{Swift:2008fm}.\\
The shock compression was driven by a $(14 \pm 1)$\,J pulse with a full duration of approximately 3\,ns, with a central wavelength of 800\,nm.
The pulse was derived from the same source as the wakefield drive pulse to minimise timing jitter.
It was focussed by an $f$/2 off-axis parabolic mirror to a point just before the shock target. 
The defocussed beam at the surface of the target had a width of approximately 1.7\,mm, chosen to be significantly larger than the silicon surface to limit lateral release. 
Its intensity profile is shown in Figure \ref{fig:shockimsetup}~(b) where it is compared to the size of the sample.
An astigmatic line focus, created by a misalignment of the parabolic focussing optic, allowed a 2D shock wave to be driven in to the target that was close to uniform in the x-ray probing direction.
It also meant that with the limited laser energy available it was possible to apply a high peak drive pressure to part of the target.
This is highlighted in Figure \ref{fig:shockimsetup}~(c) where the intensity integrated over the width of the target is shown as a function of $y$.
The peak intensity that interacted with the target was in the range $2-12 \times 10^{12}$\,\si{\watt\per\square\centi\metre}, where the range arose from the uncertainty in the $z$ position of the target, and from variations in the pointing of the drive beam. \\
Figure \ref{fig:shockprop1} shows two images of laser driven shock waves propagating in silicon taken with betatron x-rays. 
The target was placed 0.12\,m after the source implying a geometric magnification of 30.
Figure \ref{fig:shockprop1}~(a) is a radiograph of a silicon target $\Delta t = 5.2$\,ns after the start of the laser-silicon interaction. 
Figure \ref{fig:shockprop1}~(b) is an image of a silicon target with a CH ablator layer on the drive surface taken 6.5\,ns after the start of the interaction. 
Bremsstrahlung noise from electron beam interactions and shot noise was reduced in these images by performing a morphological opening with a $\left(3 \times 3 \right)$ pixel kernel, which was smaller than the system point spread function, and then applying a $\left(3 \times 3 \right)$ pixel mean filter.
The images show the optical density of the target integrated along the x-ray propagation direction, meaning that the most dense material is darkest on the image. 
From these images the position and shape of the shock front can be determined.\\
It is interesting to note that the CH ablator layer is imaged at high resolution.
This is only possible because of phase contrast effects in the image, as the ablator layer was otherwise transparent to the x-rays; the difference in detected x-ray intensity between the ablator layer and the vacuum was less than 1\%. 
The use of a plastic ablator, while affording several advantages, can complicate the interpretation of a dynamic loading experiment \cite{Swift:2008fm}. 
A thorough understanding of the dynamics of the ablator is critical to minimise experimental uncertainties.
The phase contrast from the betatron source makes it possible to study low-$Z$ ablator dynamics while simultaneously retaining the ability to probe medium-$Z$ targets. 
This has the potential to enhance our current understanding of the ablation plume and hence instruct the design of future laser driven experiments. \\
In Figure \ref{fig:shockprop1}~(c), the positions of of the shock front and the drive surface are plotted as a function of $y$ at an x-ray delay $\Delta t = 5.2$\,ns, for a target without an ablator layer.
The error in distance was governed by the resolution of the x-ray imaging system. 
The full width of the error bar, at 3.6\,\si{\um}, is the inverse of the spatial frequency for which the modulation transfer function of the imaging system was equal to 5\%, which was the approximate noise level in the image relative to the signal.
For shorter x-ray delay times, such that the drive surface had only moved a distance comparable with the point spread function of the imaging system, the relative uncertainties in its position are consequently large. 
It was critical to accurately determine the initial positions of the drive and rear surfaces of the target. 
A reference (x-ray only) image of the target provided its original thickness on the detector. The target edges were found using a Sobel edge detection algorithm. 
The original position of the drive surface in Figure \ref{fig:shockprop1}~(a) was found by finding the rear surface position with the same edge detection algorithm and subtracting the measured target width.
In this way, the effect of the pointing fluctuation of the x-ray source was removed.\\
The laser intensity is also plotted in Figure \ref{fig:shockprop1}~(c), demonstrating that its spatially varying profile drove the sample to a range of shocked states in a single experiment.
A future experiment utilising the betatron x-ray probe and a longer, temporally and spatially profiled shock drive pulse would allow direct, accurate, spatially varying measurements of shock states in a similar manner to previous single shot Hugoniot measurements \cite{Bolme:2008jl}.

\subsection*{Radiation Hydrodynamics Simulations of the Interaction}

\begin{figure}[ht]
\centering
\includegraphics[width=0.9\linewidth]{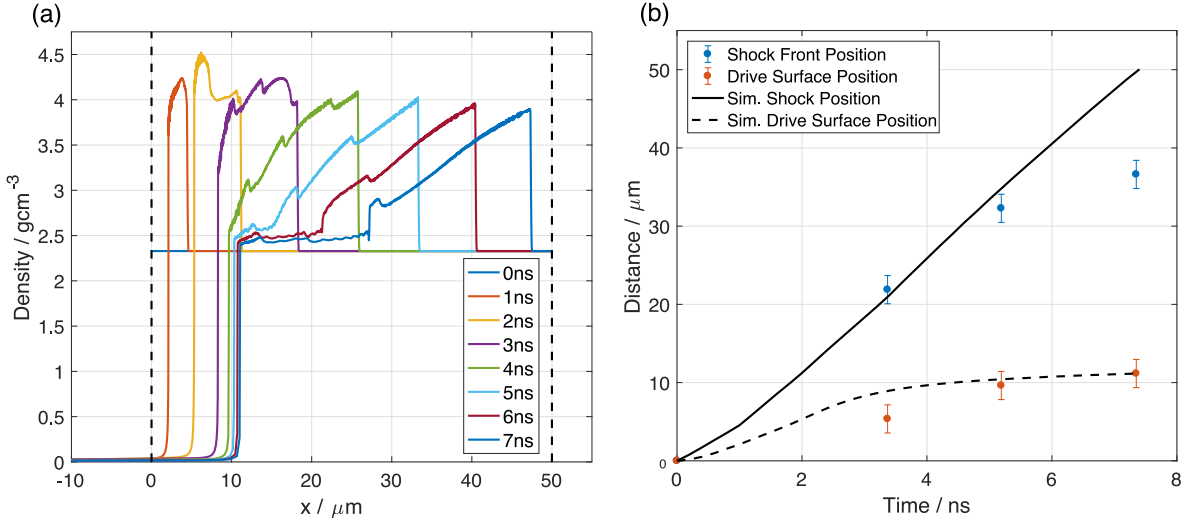}
\caption{Analysis of the shock wave propagation. 
(a)~Simulated density profiles from the 1\,D HYADES simulation of 50\,\si{\um} thick silicon irradiated by a laser pulse with a scaled peak intensity of $1.25 \times 10^{12}$\,W/cm\textsuperscript{2}, that followed the measured temporal pulse profile (see Methods). 
The black dashed lines highlight the initial edges of the target.
(b)~Shock front position (blue) and drive surface position (red) plotted from an experimental time series as a function of delay $\Delta t$ between the start of the interaction and the arrival of the x-ray pulse, at the position where the strongest shock was driven. 
This was for samples without an ablator layer, taken from 3 separate shots.
The positions of the shock and the apparent drive surface positions from the simulation are plotted as black and black dashed lines respectively.}
\label{fig:usupgraph}
\end{figure}

The measurements of the shock front and drive surface positions of the silicon sample without an ablator layer were compared to radiation-hydrodynamics simulations that were performed in the 1\,D hydrocode HYADES \cite{larsen1994179} (see Methods section for further details).
The simulated silicon density as a function of $x$ is shown at times between 0 and 7\,ns in Figure \ref{fig:usupgraph}~(a), where the driven compressive front is observed to propagate from the drive surface, initially at 0\,\si{\um}, to the right with each timestep.
The Figure shows that a structured wave was driven in to the target over the first 3\,ns, with an initial sharp shock followed by additional compressive fronts. 
This structure arises primarily from the temporal profile of the drive pulse, and to a lesser extent from the equation of state used.
After 3\,ns the drive switches off and the ablation pressure drops rapidly, causing a rarefaction wave to propagate into the compressed material, releasing the stress at the shock front. 
The position of the leading edge of the shock front and drive surface are manifest by sharp discontinuous increases, and decreases, in density from the ambient silicon density of 2.33\,\si{\gram\per\cubic\centi\metre}.
The simulated positions are compared to drive surface and shock front positions extracted from betatron radiographs captured over a range of drive to betatron delay times in Figure \ref{fig:usupgraph}~(b).
The experimental drive surface and shock positions were taken from the position where the shock was strongest i.e. at the transverse displacement at which the laser intensity was highest ($y \approx 170$\,\si{\um}).
The spatial error bars came from the resolution of the imaging system.
The error in the time, derived from the response time of the diodes used to detect the relative arrival times of the shock drive and wakefield driver lasers, was on the order of tens of picoseconds and is not shown. \\
Good agreement is observed between the experimental and simulation data at early times.
However, at later times the simulation overpredicts the shock front velocity and hence position. 
This is expected as the spatially varying drive profile will necessarily lead to lateral release and cooling effects which will result in the 1D simulations over-predicting the material compression at late times compared to the experiment.
Figure \ref{fig:usupgraph}~(b) also shows that over the first 5.2\,ns the shock velocity was approximately constant,  and that in this time the shock front had propagated $\left(32 \pm 2 \right)$\,\si{\um}, corresponding to an average $u_s = \left(6.2 \pm 0.4 \right)$\,\si{\kilo\metre\per\second}, slightly below the simulated shock velocity of 6.7\,\si{\kilo\metre\per\second}.\\

\subsection*{Measuring Density from Betatron Radiographs}

\begin{figure}[h!]
\centering
\includegraphics[width=0.5\linewidth]{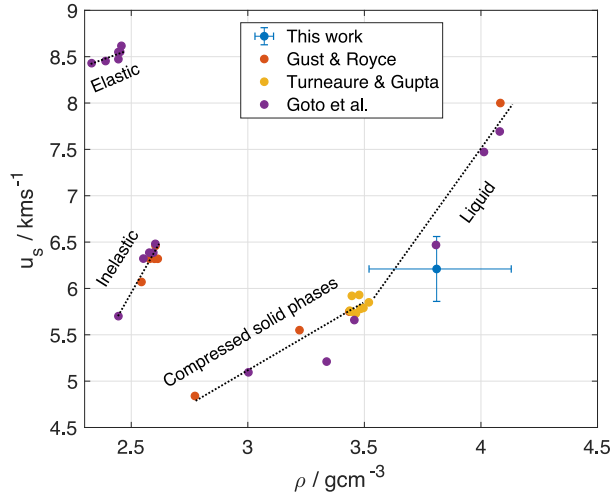}
\caption{Shock velocity versus density from the strongest part of the shock in this work compared to the data of Gust and Royce \cite{Gust1971},Turneaure and Gupta \cite{Turneaure2007} and Goto et al. \cite{Goto1982a} The black dotted lines are drawn to guide the eye to the various phases of silicon present in the graph.}
\label{fig:us_rho}
\end{figure}

An additional advantage to x-ray probing is that the density $\rho_s$ of the shocked material can be measured, provided that the x-ray spectrum is known.
In this analysis lateral release effects in the x-ray propagation direction were neglected and it was assumed that the target remained at a constant thickness $d = 200$\,\si{\um} in a compressed state of higher density. 
The x-ray transmission $T$ through the target relative to the transmission with no target in the beam $T_0$ is given by
\begin{equation}
\frac{T}{T_0} = \frac{\int_0^{\infty} \frac{\mathrm{d}I(E,E_c)}{{d}E} \ M(E) \ Q(E) e^{-\frac{\mu(E)}{\rho_0} \rho_s d} \ \mathrm{d}E}{\int_0^{\infty} \frac{\mathrm{d}I(E,E_c)}{{d}E} \ M(E) \ Q(E) \ \mathrm{d}E},
\label{eq:trans1}
\end{equation}
where $\frac{\mathrm{d}I(E,E_c)}{{d}E}$ describes the intensity of the on-axis synchrotron spectrum, $M(E)$ is the fractional transmission of other materials in the beamline such as the vacuum window and the laser block, $Q(E)$ is the quantum efficiency of the detector and $\mu(E)/\rho_0$ is the mass absorption coefficient of silicon. 
Assuming that the spectrum is well characterised by the on-axis synchrotron spectrum, the only unknown quantities in this equation are $E_c$ and the density of the shocked material $\rho_s$. 
As detailed in the Methods section, a value of $E_c = \left( 17.2 \pm 0.8 \right)$\,keV was found by comparing the detected x-ray intensity through ambient density regions of the target to the detected x-ray intensity through the region to the side of the target.
The transmission through a $17 \times 17$~pixel region just behind the shock front divided by the transmission through the region without the target was $T/T_0 = 72.1 \pm 0.2$\,\%. A least squares fitting of this value to its analytical expression gave a best fit $\rho_s = 3.81^{+0.29}_{-0.26}$\,\si{\gram\per\cubic\centi\metre}, where the uncertainties in $E_c$ and $T/T_0$ have been included.
This is plotted and compared to the silicon Hugoniot data in Figure \ref{fig:us_rho}, where lines have been drawn to divide the graph in to the various phases of silicon: elastic and inelastic responses, high pressure solid phase or phases, and finally the liquid phase \cite{Strickson2016,Gust1971,Turneaure2007,Goto1982a}.
Within errors, there is good agreement between the $u_s - \rho_s$ data measured with betatron x-ray radiation and data from the literature, which was obtained with a high explosive \cite{Gust1971} or flyer plate \cite{Turneaure2007} driver and measured with optical surface-based techniques.
The density $\rho_s$ also agrees, within error, with the peak simulated density from Figure \ref{fig:usupgraph}~(a).
It would be expected that phase contrast effects would lead to an overestimate of the density, although density measurements were taken a small distance away from phase contrast features at edges to minimise this effect. 
The assumption that the target did not expand in $z$ would have also have led to an overestimate of the density. 
Additionally, the shock velocity measured here may be expected to be lower than the Hugoniot data because of the non-ideal, ramped temporal profile of the drive pulse. 
Nevertheless, the errors introduced by these simplifying assumptions were small enough that good agreement between the experimental data and the silicon Hugoniot data was found. 

\section*{Discussion}
It has been shown that betatron radiation from a LWFA is an effective source for imaging rapidly evolving phenomena. 
Hard x-ray images of a laser driven shock wave propagating in silicon had a spatial resolution similar to what has been achieved at synchrotron light sources but with a significantly better temporal resolution limit of $<100$\,fs.
The intrinsic short pulse nature of betatron radiation ensured that motion blur of the 6.2\,\si{\kilo\metre\per\second} shock wave was negligible. 
Furthermore, we have shown that x-ray radiographs taken with the LWFA betatron source can be used to make quantitative measurements of properties of the shocked silicon target.
In this experiment the density and shock velocity were measured directly and showed a good agreement with previously published results.
The availability of a longer drive pulse to support the shock state for longer would have permitted accurate, direct measurements of the particle velocity from the motion of the drive surface of the target.
Since the error in the shock and drive surface positions was limited by the resolution of the imaging system which was constant in time, a longer shock drive pulse would have lead to a smaller relative error in the shock velocity.
The accuracy of the density measurement could be improved in future by increasing the brightness of the betatron x-ray source which would reduce the shot noise. 
Phase contrast effects, which were difficult to quantify in this experiment due to bremsstrahlung noise in the image, could be accounted for directly by simulation, or minimised by enhanced imaging spatial resolution.
In the case where a CH ablator was used the phase contrast ability of the LWFA betatron source was shown to be advantageous, allowing the deformation of the low-$Z$ ablator layer to be monitored while simultaneously probing the significantly more opaque dynamics of the silicon target.\\
This work has shown that betatron radiation has advanced to a level where it can make an impact in the field of HEDP. 
It is currently the case that high power, short pulse laser facilities are generally not equipped to meet the demands of the shock physics community, especially when it comes to providing the ideal flat-top temporal pulse profile for the shock driver that is necessary to support a shocked state for many nanoseconds.
Likewise many HEDP facilities do not have a compact source of brilliant x-rays. 
It is noted that laser driven x-ray backlighters comprising of a high energy laser pulse hitting a solid target have been implemented at some facilities \cite{Workman2008,Brambrink2009,Park2008a,LePape2010}, although these have a larger source size and therefore worse resolution, a longer pulse duration and worse collimation than betatron radiation.
Radiation produced by the LWFA mechanism has the additional advantage that it can be driven by commercially available titanium:sapphire laser systems which, because they produce $\sim 10$\,J rather than multi-kilojoule pulses, are lower footprint and higher repetition rate.
It seems likely that in the future, LWFAs installed at permanent HEDP facilities will allow a broad range of users to reap the benefits of bright, hard, pulsed synchrotron light, and to exploit the unique nature of betatron radiation: that it is femtosecond duration, high brightness and broad bandwidth.

\section*{Methods}

\begin{figure}[ht]
\centering
\includegraphics[width=0.9\linewidth]{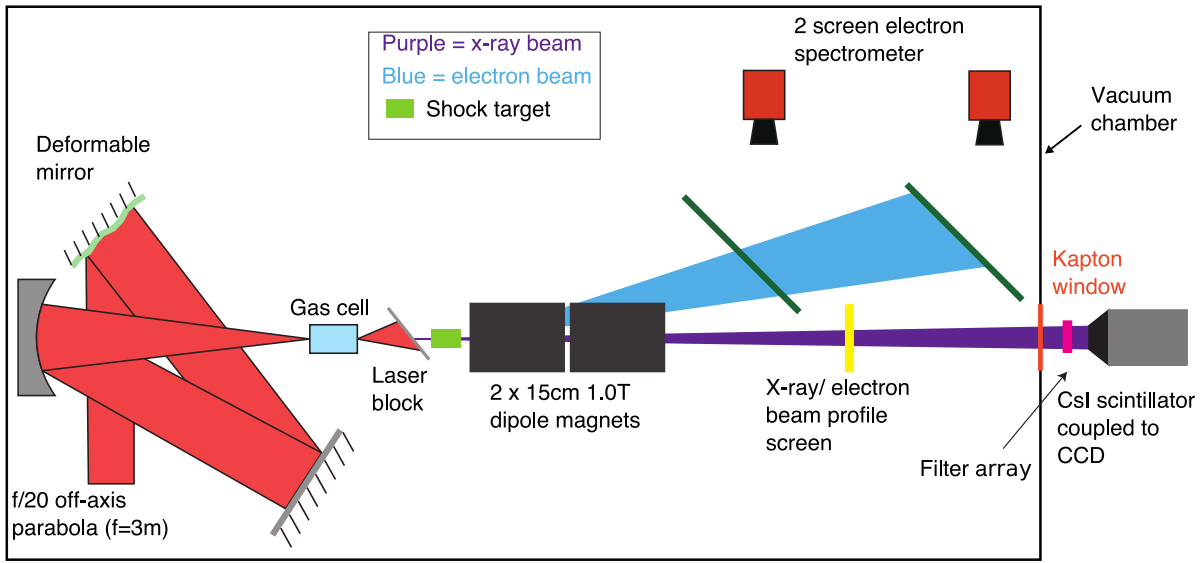}
\caption{Diagram of the LWFA betatron source showing the diagnostics used for its characterisation.}
\label{fig:setup1}
\end{figure}

\begin{figure}[h!]
\centering
\includegraphics[width=0.5\linewidth]{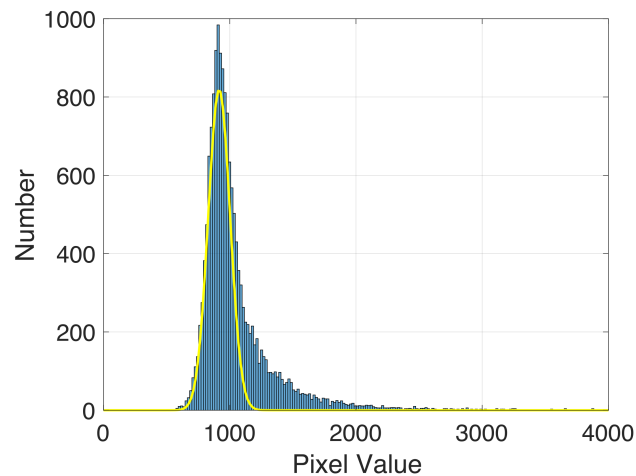}
\caption{Histogram of the pixel values in the region without the target (blue bars) overlaid with a gaussian fit to the low-count population (yellow), which is attributed to the betatron x-ray signal.}
\label{fig:bghist}
\end{figure}

\textbf{LWFA Drive Laser.}
The wakefield accelerator was driven by the Gemini South beam: a linearly polarised pulse with a central wavelength of 800\,nm, with an on-target energy of $\left(10.5 \pm 0.7 \right)$\,J and a full-width-half-maximum (FHWM) duration of approximately 50\,fs. 
The pulse, after wavefront corrections were performed with a deformable mirror, was focussed by an $f/20$ off-axis parabolic mirror to a FWHM spot size of $\left(24 \pm 5 \right) \times \left(15 \pm 1 \right)$\,\si{\um}, where the quoted error is the standard deviation.
The peak intensity at focus was $\left(2.3 \pm 0.5 \right) \times 10^{19}$\,\si{\watt\per\square\centi\metre}, corresponding to a normalised vector potential $a_0 = 0.855 \lambda [\si{\um}] I^{1/2} [10^{18}\,\textrm{W\,cm\textsuperscript{-2}}] = 3.3 \pm 0.4$. 
After the gas cell the transmitted laser energy was blocked by a 36\,\si{\um} Al foil.\\ 
\textbf{Plasma Target.} The target was a variable length gas cell with few-hundred micrometre conical apertures at the ends.
The cell length was altered by changing the separation of the tips of the cones.
Before the laser shot the gas cell was filled with neutral helium gas which was ionised by the leading foot of the laser pulse. 
The density was controlled by varying the backing pressure, and characterised on-shot by means of a transverse probe beam and a Mach-Zehnder interferometer. 
The x-ray parameters were optimal for imaging when the cell was 10\,mm long and the plasma electron number density was within the range $\left(2.1-2.8 \right) \times 10^{18}$\,cm\textsuperscript{-3}.\\ 
\textbf{Electron Beam Characterisation.} The LWFA electron and betatron x-ray beams were characterised using the set-up shown in Figure \ref{fig:setup1}.
The electron beam energy spectrum was measured by dispersing it with two 1.0\,T, 15\,cm long permanent dipole magnets on to two scintillating Gd\textsubscript{2}O\textsubscript{2}S:Tb (Kodak Lanex Regular) screens, which were imaged on to CCD cameras. 
Electrons were tracked through the measured magnetic field via particle tracking simulations so that the electron beam position on the screens could be related to their kinetic energy. 
The low energy cut-off of the diagnostic was 200\,MeV. 
The screens were calibrated using Imaging Plate \cite{Wood2016} to allow a calculation of the electron beam charge to be made.\\ 
\textbf{X-ray Beam Characterisation.} The spatial profile of the betatron x-ray beam was measured by a scintillating Gd\textsubscript{2}O\textsubscript{2}S:Tb screen placed on the laser axis imaged on to a CCD camera.
To measure the spectrum of emitted x-rays the spatial profile screen was removed from the beamline and the x-rays propagated through a 250\,\si{\um} thick kapton vacuum window and 40\,cm of air to a structured caesium iodide (CsI) scintillator that was fibre coupled to a cooled CCD camera (Princeton Instruments PIXIS-XF). 
The spectral shape was assumed to be that of the on-axis synchrotron spectrum of some critical energy that was to be determined. 
The critical energy was estimated by detecting the transmission of the x-ray beam through an array of elemental filters which was placed just in front of the CsI scintillator, which was 3.70\,m from the source.
The observed filter transmission $Y_i$ was compared to the expected transmission $T_i$, for each filter $i$, calculated for spectra with different critical energies. 
The best fit critical energy was found by minimising the quantity $\Delta(E_c) = \sum_i \left[Y_i - T_i\left(E_c \right) \right]^2$ with respect to $E_c$.
The scintillator and camera combination was calibrated with a radioactive Fe-55 source of known activity so that the number of camera counts could be converted in to energy deposited in the scintillator. 
This was converted in to a total number of photons in the beam using the measured spectral shape and beam size, since the beam took up a larger solid angle than the calibrated detector. \\
\textbf{Radiation-Hydrodynamics Simulations.}
Simulations were performed using the one-dimensional Lagrangian radiation hydrocode HYADES \cite{larsen1994179}. 
Silicon was modelled without strength, and using a wide-ranging QEOS (table 392). Opacities for radiation diffusion were taken from the SESAME database, and a Thomas-Fermi ionization model was used.  
The 50\,\si{\um} silicon wafer was discretised into approximately 1800 zones, feathered towards the drive surface. The 1\,D simulations were performed using the peak measured irradiance history, but were scaled by 66\% in an effort to account for the reduction in ablation pressure arising from radial heat transport \cite{Reighard:2007kk}. 
An apparent peak irradiance of $1.9 \times 10^{12}$\,\si{\watt\per\square\centi\metre} (scaled to $1.25 \times 10^{12}$\,\si{\watt\per\square\centi\metre}) provided the best qualitative match to the observed shock and drive surface motion, comparable to, albeit slightly below, the measured peak irradiance of ($2-12 \times 10^{12}$\,\si{\watt\per\square\centi\metre}), which was known to have a large uncertainty.
Additionally, it should be acknowledged that the non-uniform drive profile (Figure \ref{fig:shockimsetup}~(b)) will necessarily lead to lateral release and cooling effects which will result in the 1D simulations over-predicting the material compression at late times.\\
\textbf{Density Estimation.}
The x-ray transmission $T_a$ through ambient density regions of the silicon target, relative to the intensity in the absence of the target $T_0$ can be expressed by Equation \ref{eq:trans1} with the substitution $\rho_s \rightarrow \rho_0$.
The value of $T_a/T_0$ was found from the experimental x-ray image by dividing the x-ray intensity in regions of ambient density silicon close to the shock front by the x-ray intensity in a nearby (background) region to the right of the shock target.
A least squares fitting was applied to Equation \ref{eq:trans1} to determine the best fit $E_c$.
To accurately measure $T_a/T_0$ it was critical to apply a method of noise removal that left most of the pixel values unchanged.
First the camera dark field, i.e.~the image recorded when the x-ray source was switched off, was subtracted from the image.
The image was then divided by the camera flat field, which is the response of the x-ray camera to uniform illumination, to correct for the variation in pixel response.
A histogram of the pixel values in the background region is shown in Figure \ref{fig:bghist}.
The shape of this histogram can be explained by a low pixel value population, which is attributed to the `true' betatron signal, and a long tail of high pixel values which is caused by hard x-ray noise, primarily from bremsstrahlung x-rays generated by electron beam interactions.
A two gaussian fit was made to the shape of this histogram. 
As shown in Figure \ref{fig:bghist} one of these gaussians was a good fit to the low pixel value population.
Its mean value was taken as the betatron x-ray intensity in this region, and the standard error on the mean was calculated from the width of the distribution and the number of pixels it contained.
This method was applied to find both $T_a$ and $T_0$, from which $E_c$ was estimated.
The error in $E_c$ was propagated from the standard error on $T_a$ and $T_0$.
The transmission $T$, expressed in Equation \ref{eq:trans1}, through a $17 \times 17$\,pixel region just behind the shock front was found by the same method, and with the measured value of $E_c$ it was possible to calculate the density of shocked material $\rho_s$ via a least squared fitting to $T/T_0$.

\footnotesize
\bibliography{references2.bib}

\section*{Acknowledgements}

The authors thank the laser, engineering and target fabrication staff at the Central Laser Facility, Rutherford Appleton Laboratory for their hard work. 
This research was supported by STFC (ST/J002062/1, ST/P000835/1),  EPSRC (EP/I014462/1), and the European Research Council (ERC) under the European Union's Horizon 2020 research and innovation programme (Grant Agreement No. 682399). 
The Institute of Shock Physics acknowledges the support of AWE and Imperial College London. 
Part of this work (F.A. and B.B.P) was performed under the auspices of the U.S. Department of Energy by Lawrence Livermore National Laboratory under contract DE-AC52-07NA27344, supported by the LLNL LDRD program under tracking code 13-LW-076. F.A. acknowledges support from the DOE Office of Science Early Career Research Program under SCW 1575.
K.T.B., K.K., A.G.R.T. and Z.Z. acknowledge funding from US DOE under grant DE-NA0002372.
The work of W.S. and S.H.G. was supported by US DOE Fusion Energy Sciences under FWP 100182.

\section*{Author contributions statement}

The experiment was conceived by S.P.D.M., N.B., S.G., Z.N., B.B.P., D.E., S.R., M.S., F.A. and E.H. 
The experiment was conducted by J.C.W., D.J.C., K.P., N.C.L., M.E.R., T.W., J.S.J.B., P.F. and W.S., along with K.T.B. and Z.Z. who were supervised by K.K. and A.G.R.T. 
The data was analysed and the manuscript was written by J.C.W., and it was improved through discussions with D.J.C, M.E.R., Z.N., S.P.D.M. and D.E.
The simulations were performed by D.J.C. with valuable input from R.S.

\section*{Additional information}

No authors declared any competing financial interests. 
The authors confirm that all of the data used in this study are available without restriction. Data can be obtained by contacting plasma@imperial.ac.uk.



\end{document}